\documentclass[sigconf]{acmart}
\AtBeginDocument{%
  }

\setcopyright{acmlicensed}
\copyrightyear{2018}
\acmYear{2018}
\acmDOI{XXXXXXX.XXXXXXX}
\acmConference[Conference acronym 'XX]{Make sure to enter the correct
  conference title from your rights confirmation email}{June 03--05,
  2018}{Woodstock, NY}
\acmISBN{978-1-4503-XXXX-X/2018/06}

\usepackage{multirow}
\usepackage{enumitem}
\usepackage[table]{xcolor}

\begin{document}

\title{An Index-based Approach for Efficient and Effective Web Content Extraction}



\author{Yihan Chen}
\affiliation{%
  \institution{University of Science and Technology of China}
  \city{Hefei}
  \country{China}}
\email{chenyihan@mail.ustc.edu.cn}

\author{Benfeng Xu}
\affiliation{%
  \institution{University of Science and Technology of China}
  \city{Hefei}
  \country{China}}
\email{benfeng@mail.ustc.edu.cn}

\author{Xiaorui Wang}
\affiliation{%
 \institution{Metastone Technology}
 \city{Beijing}
 \country{China}}

\author{Zhendong Mao}
\affiliation{%
  \institution{University of Science and Technology of China}
  \city{Hefei}
  \country{China}}

\renewcommand{\shortauthors}{Trovato et al.}

\begin{abstract}
As web agents (e.g., Deep Research) routinely consume massive volumes of web pages to gather and analyze information, LLM context management—under large token budgets and low signal density—emerges as a foundational, high-importance, and technically challenging problem for agentic and RAG pipelines. Existing solutions for extracting relevant content are inadequate: generative extraction models suffer from high latency, rule-based heuristics lack adaptability, and chunk-and-rerank methods are blind to webpage structure.
To overcome these issues, we introduce \textbf{Index-based Web Content Extraction} to reframe the extraction process from slow, token-by-token generation into a highly efficient, discriminative task of index prediction, achieving both effectiveness and efficiency. We partition HTML into structure-aware, addressable segments, and extract only the positional indices of content relevant to a given query. This method decouples extraction latency from content length, enabling rapid, query-relevant extraction.
We first evaluate our method as a post-retrieval processing component within an RAG QA system and find that it improves QA accuracy. Then we directly measure its match rate with the target content in two scenarios: main content extraction (ME) and query-relevant extraction (QE).
Experimental results show that our method outperforms existing works in both accuracy and speed, effectively bridging the gap between LLMs and the vast webpages.
\end{abstract}



\received{20 February 2007}
\received[revised]{12 March 2009}
\received[accepted]{5 June 2009}

\maketitle

\begin{figure}[h]
\centering
\includegraphics[width=1\columnwidth]{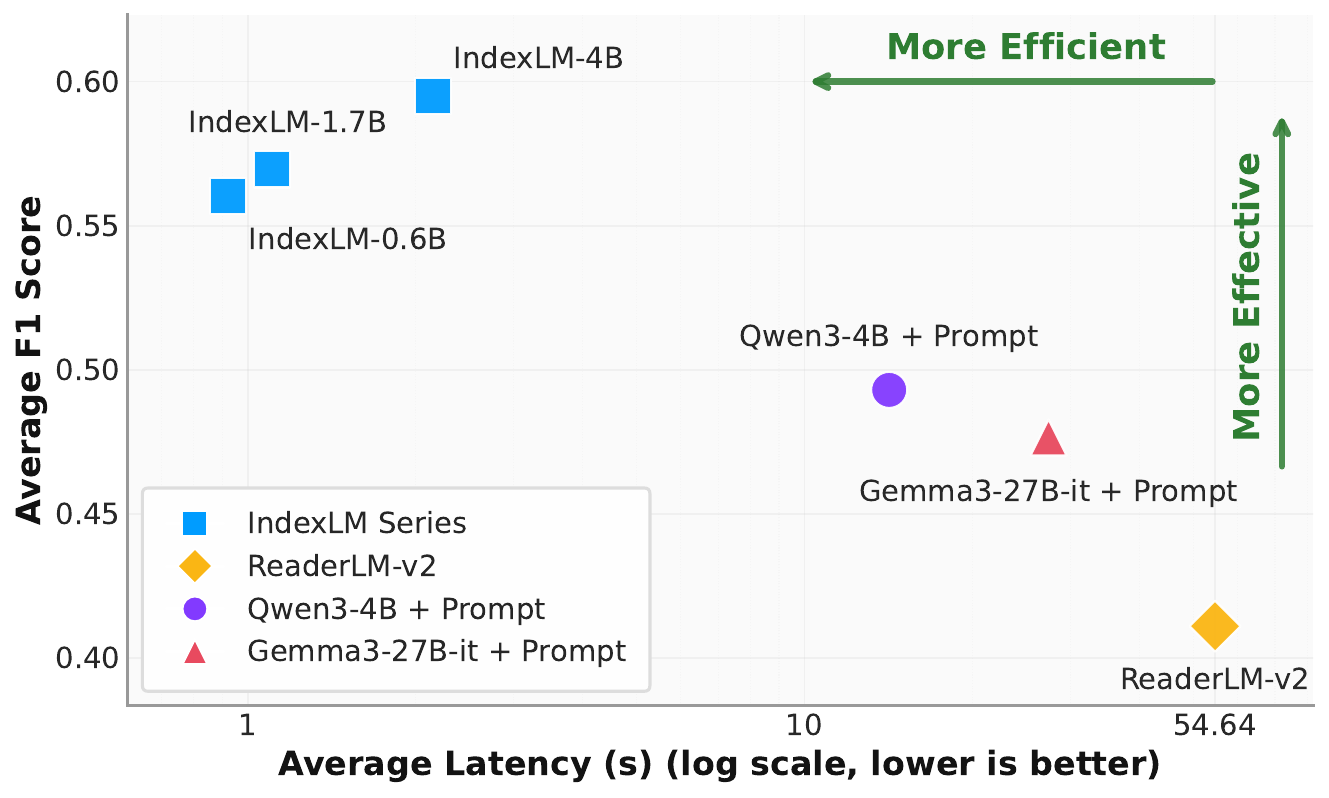}
\caption{Index-based extraction offers faster speed compared to token-by-token generative extraction.}
\label{pace}
\end{figure}

\begin{figure*}[htbp]
\centering
\includegraphics[width=\linewidth]{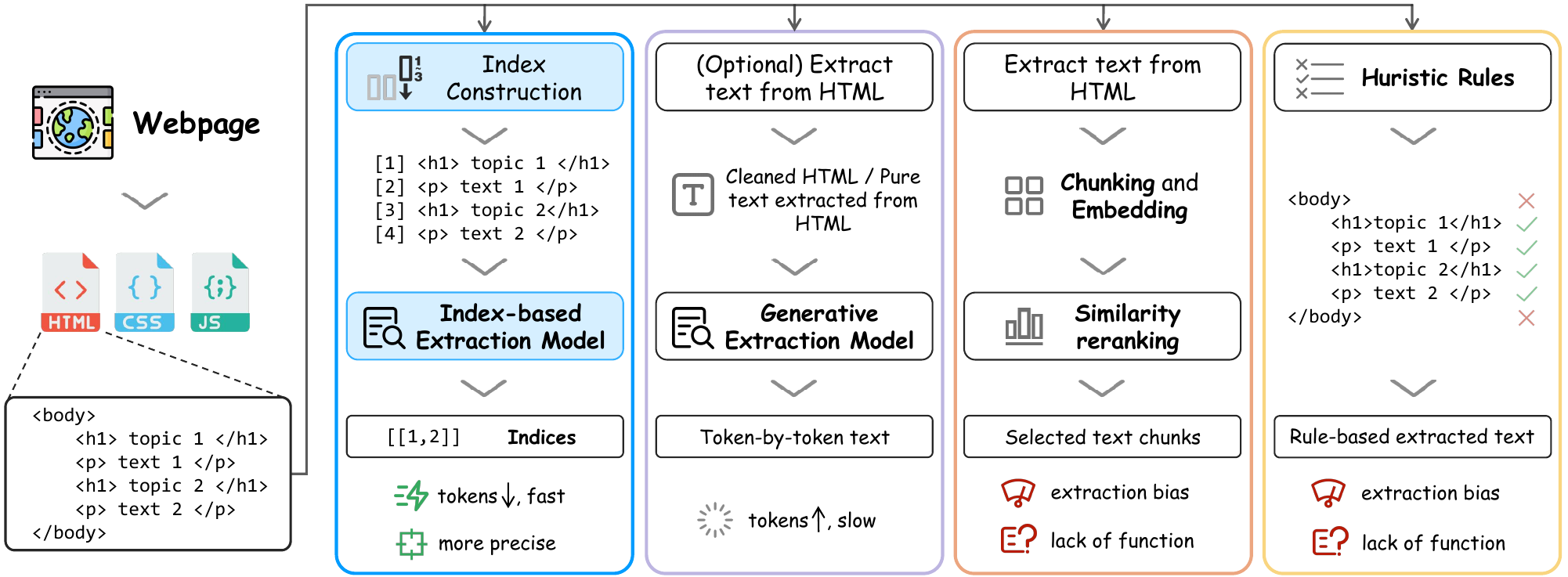}
\caption{Comparison of index-based web content extraction and previous works. Chunk and rerank RAG methods are unable to perform main content extraction, while methods based on heuristic rules have difficulty with query-relevant extraction. In comparison to other methods, our approach is both effective and efficient.}
\label{comp}
\end{figure*}

\section{Introduction}
With the continuous advancement of their capabilities, Large Language Models (LLMs) are increasingly envisioned as the foundation for autonomous web agents designed to tackle complex, real-world tasks \cite{li2025websailor, ning2025webagentsurvey}, such as answering multi-hop questions that require multiple browsing steps (e.g., GAIA \cite{zavras2025gaia}, BrowseComp \cite{wei2025browsecompsimplechallengingbenchmark}) and performing Deep Research~\cite{deepresearch, li2025webthinker, zhang2025deepresearchsurvey} that demands searching, reading, and synthesizing information from potentially hundreds of websites.
However, according to the HTTP Archive\footnote{https://httparchive.org/reports/page-weight}, as of 2025-08, the median size of a webpage’s HTML source (including CSS and JS) is 869.8 KB, roughly 890K characters and 223K tokens. It is far exceeding the 32K or 128K context limits of most current LLMs. Even after excluding about 96\% of JS and CSS bytes, the remaining HTML still contains approximately 9K tokens, which is prohibitive for web-agent workloads that must read dozens or even hundreds of pages, and longer contexts may also cause "lost in the middle" \cite{lostinthemiddle}. Moreover, the token count contained in HTML continues to trend upward.
Therefore, \textbf{there is an urgent need for an efficient extraction model that bridges LLMs and the vast amount of webpages} to extract only the information truly relevant to the task, providing higher‑quality context for LLMs.

Researchers commonly utilize the following methods to extract content from webpages before feeding it into the LLM's context.
\textbf{(1) Model-based extraction}. Using an extraction model, including prompting an LLM to perform the extraction itself \cite{JinaAI2024ReaderLM, wang2025readerlmv2smalllanguagemodel, huang2025deepresearchagents}. However, these models are typically generative, meaning they output all parts of the extracted content sequentially. This leads to low efficiency, as the extraction speed decreases with the amount of target content. 
\textbf{(2) Heuristic method}. Using rules to extract content from HTML directly \cite{gupta2003dom, sun2011dom, insa2013dom}. However, due to the varied structures of webpages, this method cannot guarantee the complete removal of noise information, nor can it achieve query-relevant content extraction.
\textbf{(3) Chunk and rerank}. Using traditional RAG (Retrieval-Augmented Generation) methods \cite{fan2024ragsurvey, gao2024ragsurvey}, which involve segmenting the original content and using BM25 \cite{bm25, jiang2023active, chen2017reading} or embedding similarity \cite{chen2024bge, karpukhin2020dense, jin2025searchr} calculations to select relevant text chunks. However, these methods cannot understand the structural information of a webpage (e.g., which text blocks are the main body rather than navigation bars or ads). This can lead to the erroneous discarding of highly relevant text or the retention of irrelevant text, and it is also unsuitable for extracting the main content of a webpage. Furthermore, similarity calculation has its limitations, as high similarity does not equate to actual relevance.

To address the aforementioned issues, we propose a different ``\textit{Index-based Localization}" extraction paradigm. Instead of generating content token by token, we index the content, thereby enabling the extraction model to directly pinpoint the required information in the webpage and output only positional indices. 
By leveraging the intrinsic structure of HTML (e.g., div, p, li), we can partition the content along its natural boundaries. This allows us to subsequently assign indices to each segment, yielding an addressable and indexable representation of the webpage.
In contrast to prior approaches, the new paradigm offers the following advantages: \textbf{(1)} It achieves fast and stable extraction speed. Unlike generative methods, this discriminative approach of outputting indices decouples the extraction time from the length of the content to be extracted, resulting in lower latency. \textbf{(2)} It can perform query-relevant extraction and, compared with heuristic-based methods, is more flexible and versatile. \textbf{(3)} It is structure-aware: rather than relying solely on similarity, it can distinguish main content from navigation or advertisements and thereby mitigate misses and noise. 

Therefore, building upon our proposed paradigm, we introduce \textbf{Index-based Web Content Extraction} to enable highly efficient and effective retrieval of information from webpages. As Figure \ref{comp} visually demonstrates, our method offers distinct advantages over previous approaches. To empower this framework, we also train a corresponding extraction model, named \textit{``IndexLM''}. Upon receiving the HTML source code of a webpage, we first automatically remove formatting noise and partition the content into blocks based on HTML tags, assigning each block a one-dimensional numeric index. Subsequently, given a user query, the IndexLM identifies all relevant block indices. The content from these selected indices is then reassembled into HTML format, which can be readily converted to plain text or Markdown to further reduce the token count. Our approach reframes the extraction process from "content generation" to "index prediction", enabling highly efficient extraction without sacrificing accuracy. This allows RAG or web agents to obtain more precise and contextually relevant information.

We validate the efficacy of our extraction method and IndexLM through two primary evaluation scenarios.
Recognizing that our method can enhance the performance of RAG systems, the first is integrating the extraction model as a post-retrieval processing component within an RAG QA system. We conducted experiments on five QA datasets, where the extraction model extracts relevant content from webpages based on a query. 
The second scenario directly measures methods by calculating the match rate between its output and the target content. We constructed our test sets based on the labeled val sets of HotpotQA \cite{yang2018hotpotqa} and Musique \cite{trivedi-etal-2022-musique}. Additionally, the experiments mentioned above are all query-relevant extraction (QE). When no specific query is provided, we set the default behavior of the model to webpage main content extraction (ME). We constructed the main content extraction test set using manual annotation. Our extensive experiments across various settings demonstrate that index-based extraction outperforms previous methods in both accuracy and speed.
Our primary contributions are as follows:
\begin{itemize}[leftmargin=*]
\item We propose Index-based Web Content Extraction method to improve efficiency while ensuring extraction effectiveness.
\item We train `IndexLM' specifically for index-based web content extraction, which further improved the extraction performance.
\item We compile or construct datasets for testing extraction methods from different perspectives and conduct a comprehensive evaluation of existing extraction methods.
\end{itemize}


\section{Related Works}
\subsection{Web Content Extraction}
Early work on web content extraction focused on identifying the main content of a webpage \cite{empiricalcompwebextraction}. Such approaches generally utilize heuristic rules to identify the blocks of main content~\cite{2011removing, sun2011dom, insa2013dom} or employ machine learning methods to classify the different regions of a webpage~\cite{zhang2021boilerplate, leonhardt2020boilerplate}.
In parallel, there are also methods tailored to extracting the text span corresponding to a given target field for structured information~\cite{wang2022webformer, aveqa}.
These traditional approaches are limited either by the capacity of their underlying models or by their reliance on fixed extraction schemas, rendering them insufficiently flexible and unsuited to contemporary RAG and web-agent systems.

Apart from employing naive RAG pipelines that chunk text and perform similarity-based retrieval \cite{jin2025searchr, R1-searcher}, contemporary web-agent frameworks typically incorporate a dedicated agent step for extracting key information.
This step can be accomplished directly by the agent's backbone LLM \cite{zheng2025deepresearcherscalingdeepresearch}, or by specialized webpage-extraction models such as ReaderLM \cite{wang2025readerlmv2smalllanguagemodel}.
However, these approaches are generative and thus comparatively time-consuming. In contrast, our proposed index-based extraction is a discriminative method that achieves effectiveness while remaining efficient. 
In addition, HtmlRAG \cite{htmlrag} scores each text block in the HTML tree using embedding-model similarity for pruning and then performs further extraction with a trained generative model, iteratively removing low-scoring chunks until the remaining context fits within a target window.
However, the length of the extracted content is predetermined and not adaptive, which hinders precise extraction and can lead to either under-extraction or the retention of irrelevant information.

\subsection{Post-Processing of Retrieval}
The retrieval phase has become an indispensable component of contemporary RAG systems and web agents. However, the information returned by retrieval unavoidably introduces noise, redundancy, and irrelevant content. Consequently, it is common to incorporate a post-retrieval processing stage.
Within this stage, one approach is to rerank the chunked text segments by relevance~\cite{search-o1, moniz2024realm}. Some studies further perform reasoning before reranking and adopt list-wise reranking strategies to achieve greater speed and accuracy~\cite{list-aware-reranking, yang2025rankk, liu2025listconranker}.
Our index-based extraction serves the same purpose of context refinement by operating as an intra-retrieval filtering stage. It can be synergistically combined with other post-processing techniques. For instance, when a task involves multiple webpages, it can be paired with reranking to order the extracted content from each page, thereby further optimizing the context.

\section{Methodlogy}

\subsection{Problem Definition}
The goal of web content extraction is to extract a subset of content from a webpage that is highly relevant to a given task. That is, given a webpage represented by its raw HTML source code $H$ and a natural language user query $Q$, the extraction model $E$ will extract the content $R$ related to $Q$ from $H$.
This process can be formally represented as:
$R = E(H, Q)$.
Specifically, when $Q = \emptyset$, $E$ will extract the main content of the webpage by filtering out extraneous sections such as advertisements, navigation bars, and related articles, such that $R = \text{main\_content}(H)$.

Extraction models are typically employed in the Retrieval Augmented Generation (RAG) pipeline for LLMs. 
RAG is a framework that enhances the capabilities of LLM by retrieving relevant information from external knowledge sources to generate more accurate and up-to-date responses \cite{fan2024ragsurvey, gao2024ragsurvey}. Consequently, in many scenarios, RAG systems are required to retrieve information from the web \cite{htmlrag, zheng2025deepresearcherscalingdeepresearch, jin2025searchr}.
With the introduction of an extraction model $E$, the RAG workflow can be represented by the following pipeline:
$$Q_{\text{user}} \xrightarrow{\text{Retrieve}} \{H_1, H_2, \dots\} \xrightarrow{E} \{R_1, R_2, \dots\} \xrightarrow{\text{LLM}} A_{\text{final}}$$
The RAG system first retrieves a set of raw webpages ${H_i}$ based on a user query $Q_{\text{user}}$. The extraction model $E$ then processes each page to produce a set of more concise and query-relevant content $\{R_i\}$. This refined content, together with the original query, is finally used by the LLM to generate the answer $A_{\text{final}}$.

\begin{figure*}[htbp]
\centering
\includegraphics[width=0.98\linewidth]{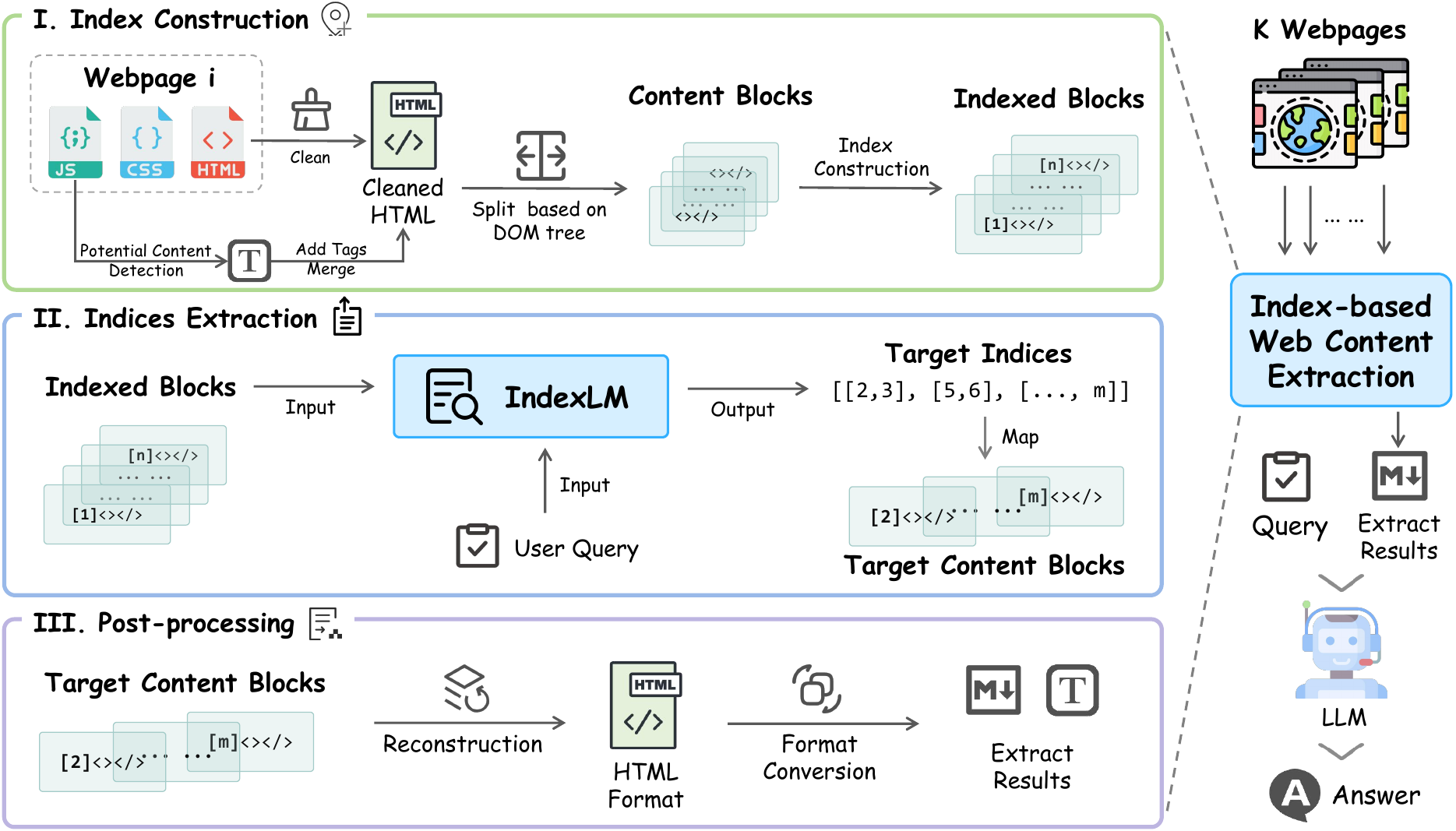}
\caption{The complete process of Index-based Web Content Extraction.}
\label{framework}
\end{figure*}

\subsection{Index-based Web Content Extraction} \label{ie}
\begin{figure*}[htbp]
\centering
\includegraphics[width=0.98\linewidth]{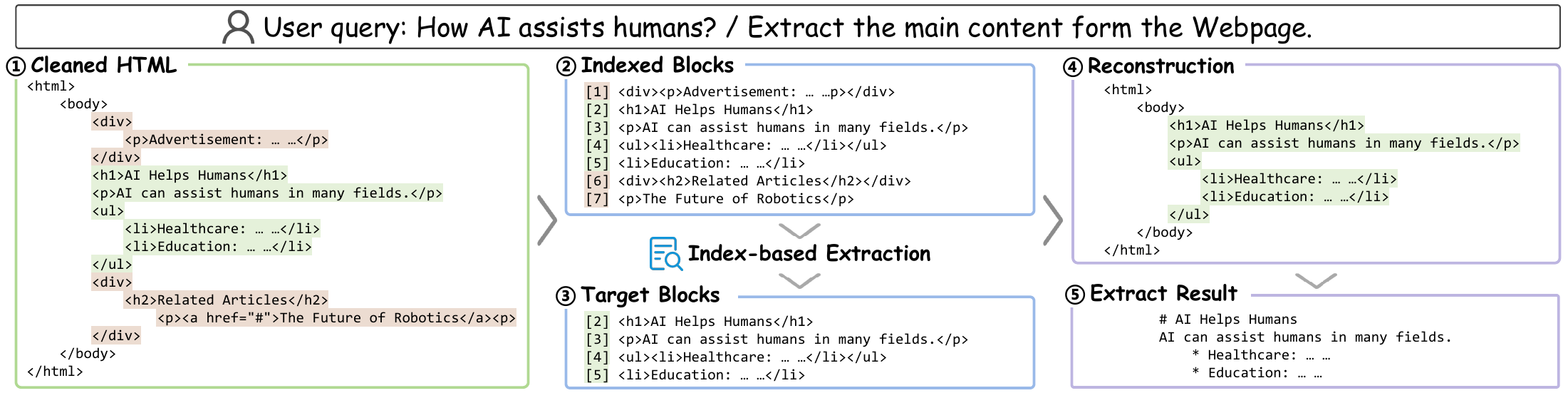}
\caption{An example of Index-based Web Content Extraction. Text with a green background represents query-relevant content, while red indicates the opposite. The original content will be mapped to an index, and finally, the query-relevant index will be mapped back to the original text blocks. A more detailed real example is shown in Appendix \ref{example}.}
\label{example1}
\end{figure*}

In this paper, we propose Index-based Web Content Extraction, with the framework and an example presented in Figure \ref{framework} and an \ref{example1}.

The input to the system is still $H$ and $Q$. The HTML source $H$is first cleaned and then transformed into an ordered sequence of $n$ addressable content blocks, denoted as $B = \{b_1, b_2, \dots, b_n\}$. Each block $b_i$ is associated with a unique 1-dimensional numeric index $i$ and retains its original HTML tag structure (e.g., `<p>', `<h1>'). In the actual representation, the index $i$ is prepended to its corresponding block $b_i$ as a textual label. This results in a format where each block is presented as a line of text with a line number, such as `[$i$] <tag>contents</tag>'.

We train a specialized model (\textit{IndexLM}) to identify all relevant intervals. IndexLM will learn a mapping function $f$ that takes the sequence of blocks $B$ and the query $Q$ as input, and outputs a subset of indices $I^* \subseteq \{1, 2, \dots, n\}$. This process can be represented as:
\begin{equation}
    I^* = f(B, Q)
\end{equation}
These indices correspond to the set of blocks $C = \{b_i \mid i \in I^*\}$, which collectively fulfill the task defined by $Q$. The ultimate objective of the model is to maximize the relevance between the selected blocks $C$ and the query $Q$, while minimizing the inclusion of irrelevant information. The set of query-relevant content blocks $C$ is then reassembled into a new HTML document, which is subsequently converted into suitable formats.

Special Case (Main content extraction):
When the query $Q$ is not provided (i.e., $Q = \emptyset$), the task defaults to main content extraction. In this scenario, IndexLM identifies the set of indices $I^*$ that corresponds to the primary content of the webpage.

\subsubsection{Index Construction} \label{index}
\paragraph{HTML Cleaning}
We employ the BeautifulSoup\footnote{\texttt{https://pypi.org/project/beautifulsoup4/}} to parse the HTML, with the initial step of extracting the webpage title from the HTML head. Subsequently, considering the presence of non-content elements (e.g., CSS, JavaScript, Comments) within the HTML body, we proceed with an HTML cleaning process. However, we observed that a minority of webpages embed their effective content within JavaScript, while previous work typically removed these scripts during the initial cleaning phase \cite{wang2025readerlmv2smalllanguagemodel}, resulting in information loss. Therefore, our approach is to first detect and extract potential textual information (e.g., HTML strings contained within JavaScript scripts) prior to their removal. These extracted texts will be placed at the very end of all the extracted content.

\paragraph{Content Segmentation and Indexing}
To segment the cleaned HTML content, we perform a Depth-First Search (DFS) traversal of the DOM tree. The traversal preferentially begins at the body node. If the body is absent, the outermost nodes are sequentially considered as alternative starting points. 

In the case of block-level elements, each element is typically mapped to a single segment. Elements that contain neither textual content nor child elements are disregarded during processing. 
\textbf{(1) }If a parent element contains direct text (text not encapsulated within a child block-level element), the parent’s textual content and its inline element text are consolidated into a single segment; each child block-level element is then recursively processed, with its resulting segments appended in sequential order.
\textbf{(2) }If a parent element lacks direct text, it merges with its first child block-level element to form a single segment (e.g., \texttt{<div><p>text</p></div>}).

In the case of inline-level elements, none of them are treated as individual segments, except for images.
\textbf{(1)} Images are preserved only if they are associated with a textual caption. The image links and captions will be formatted into a new format.
\textbf{(2)} For other inline elements, we preserve tags that denote formatting (e.g. \texttt{<b>}, \texttt{<i>}, \texttt{<br>}), but strip the tags from hyperlinks (<a>), keeping only the anchor text. All other inline tags are removed, retaining only their text content. The original ordering of all inline content is maintained throughout processing.

Given that textual content within an element can be excessively long, we define a maximum block length limit $L_{max, block}$. To avoid disrupting the text's structure, any block exceeding this limit will be split. We add a sequence-marking attribute to the tags of its blocks to facilitate subsequent reassembly. More details on content segmentation can be found in Appendix \ref{method_details}.

\paragraph{Adding Index Tags}
After the aforementioned operations, we obtain $n$ addressable content blocks, denoted as $B = {b_1, b_2, \dots, b_n}$. A unique numerical index $i$ is then assigned to each block $b_i$, in sequential order from the beginning to the end.

\subsubsection{Indices Extraction}
The extraction model $E$ generates a query-relevant index set $I^*$ based on the query $Q$, the webpage title, and the index-tagged webpage blocks. The index set is represented as a set of closed intervals, for example, `[1,1], [3,5]'. If no relevant content exists on the webpage, $E$ will output 'NA'. It then retrieves the content $C = \{b_i \mid i \in I^*\}$ from the blocks $B$.

Due to the extraction model's limited context window, we introduce a maximum document token length limit $L_{max, doc}$. The collection of blocks $B$ is partitioned into chunks, ensuring that the total number of tokens in each chunk does not exceed $L_{max, doc}$. The model $E$ runs the extraction process on each chunk of $B$ separately, outputting a series of index sets. Finally, these index intervals are merged to produce the final result. In theory, this method can process data of arbitrary web content length.

\subsubsection{Post-processing} \label{postprocessing}
The extracted blocks $C$ are then reassembled into HTML. 
First, content blocks that were split from a single element due to length constraints are concatenated based on their attribute triplets. 
Given the complexity and diversity of the HTML structure, as well as potential losses during the extraction process, we design a heuristic method to restore the parent-child structure in the DOM tree that is common and may affect the text formatting. The remaining blocks are left unchanged, preserving their original order. More details of post-processing and this heuristic method can be found in Appendix \ref{method_details}.
After removing all index tags, these reconstructed elements are sequentially inserted into an empty `<body>' tag and merged with the head constructed using the original webpage's title information, resulting in a new HTML.

Theoretically, the final extracted text, after being reassembled into an HTML structure, can have various final conversion methods. Here, we chose Markdown which is commonly used \cite{JinaAI2024ReaderLM} as the primary conversion strategy, utilizing Markitdown\footnote{https://github.com/microsoft/markitdown} to accomplish this task. The Markdown format preserves essential formatting while simplifying the content structure for subsequent tasks.

\subsection{IndexLM for Index-based Extraction}
\subsubsection{Datasets} \label{traindata}
The training data consists of two components: query-relevant data, which corresponds to extraction based on a specific user query $Q$, and main content extraction data, which focuses on extracting the main content from a webpage.

\textbf{Query-relevant Data.} We begin by randomly sampling 1,000 queries from each of the training sets of HotpotQA \cite{yang2018hotpotqa} and Natural Questions (NQ) \cite{kwiatkowski2019nq}. For each query $q$, we employ an LLM for query decomposition, breaking it down into a set of sub-queries $\{q_1, \dots, q_k\}$. We then use the Google Search API to retrieve 15 webpages for each sub-query, adding them to the webpage collection $W$ for $q$. To mitigate the issue of irrelevant search results, we leverage the ground-truth labels in the original datasets to identify and include the corresponding Wikipedia pages in $W$, followed by a deduplication process. Furthermore, since both HotpotQA and NQ are Wiki-based, we also incorporate the MultiHopRAG~\cite{tang2024multihoprag} dataset, which is derived from the news domain and already provides related URLs for each query. A portion of this dataset is integrated into our training set, while a smaller subset is held out for the final test set. This process results in a diverse set of query-webpage pairs.

Subsequently, we crawl the HTML source code for these webpages. Pages from which HTML cannot be retrieved are discarded. If all webpages associated with a query are inaccessible, the query itself is also removed. After tagging the HTML of each webpage with indices as described in Section~\ref{index}, we use a strong LLM to label the index intervals relevant to the query. This annotation process is repeated five times for each page, and we apply a majority vote: An index is included in the final label intervals only if it is identified in at least three of the five runs.

\textbf{Main Content Data.}
A portion of the webpages is sourced from Query-relevant Data, while another portion is randomly sampled from the Common Crawl top 500 domains\footnote{https://commoncrawl.org/blog/common-crawl-url-index}. The Common Crawl .warc files already include the HTML source. The annotation process is similar to that of constructing the Query-relevant Data, except that the annotation task is changed to extracting the main content of the webpage.

The LLM used for the aforementioned decomposition and annotation tasks is DeepSeek V3.1 \cite{deepseekai2025deepseekv3technicalreport}. To prevent an over-representation of instances with empty labels (i.e., cases where no query-relevant content is found) in the final training set $\mathcal{D}$, we perform a partial filtering of such instances.

\subsubsection{Training}
Since webpage content extraction tasks are time-sensitive, considering the computational cost, and given that smaller-scale LLMs now possess certain reasoning and instruction-following capabilities, we build our Index-based Extraction Model $E$ based on Qwen3-4B, 1.7B, and 0.6B \cite{yang2025qwen3technicalreport}. The final trained models are named as IndexLM-0.6/1.7/4B.

The training process is a typical SFT (Supervised Fine-Tuning) process \cite{zhang2025sftsurvey, qin2025unleashingsurvey}, where the input consists of a prompt $P$ containing the webpage URL, title, and content blocks with added index tags $B = \{b_1, b_2, \dots, b_n\}$. The label consists of intervals formed by index numbers, formatted as a Python list, e.g., `[[1,2],[3,5],[7,7]]', which is then converted into a string. During actual usage, a parsing tool will be added to convert it into a true Python list.

The SFT objective function can be expressed as follows, where $I$ represents the string obtained by converting the index intervals:
$$
\mathcal{L}_{SFT}(\theta) = -E_{(P, I) \sim \mathcal{D}} \left[ \log E_{\theta} (I | P) \right] \nonumber
$$

\begin{table}[t]
    \centering
    \begin{tabular}{l|ccc}
    \hline\hline
\textbf{Dataset} & \textbf{\#Queries} & \textbf{Avg. Webpage} & \textbf{Avg. Tokens} \\ \midrule
HotPotQA & 400 & 29.75 & 75498 \\
Musique & 400 & 33.00 & 69714  \\
TQ & 400 & 23.18 & 67374  \\
NQ & 400 & 25.61 & 83312  \\
MultiHopRAG & 154 & 2.72 & 118277  \\
    \hline\hline
    \end{tabular}
    \caption{QA test dataset statistics. ``Avg. Webpage" denotes the mean number of webpages the extraction model must consult to answer a dataset query. ``Avg. Tokens" is the mean token count per webpage across the corpus.}
    \label{stat}
\end{table}

\begin{table*}[htbp]
    \centering
    \definecolor{lightgray}{gray}{0.96}
    \begin{tabular}{c|l|cccccc>{\columncolor{lightgray}}c}
    \hline\hline
    \textbf{answer model} & \textbf{Extractor Method} & \textit{Avg. Tokens} & \textbf{HotpotQA} & \textbf{NQ} & \textbf{TQ} & \textbf{Musique} & \textbf{MultiHopRAG} & \textbf{Average} \\ \midrule
    \multirow{10}{*}{\textit{Qwen3-4B}}  
    & HTML (raw) & 4096 & 19.22 &26.32&64.25&5.43&63.29&35.70 \\
    & Markdown (raw) & 4096 & 29.82 & 37.74 & 72.56 & 6.71 & 68.42 & 43.05 \\
    & Chunk-Rerank & 4094 & 33.96 & 44.39 & 79.52 & 7.09 & 82.23 & 49.44 \\
    & HtmlRAG & 3562 & 32.82 & 42.56 & 78.97 & 10.01 & 70.63 & 47.00 \\
    & Qwen3-4B + prompt & 2563 &39.23&45.23&84.43&\textbf{12.64}&76.88&51.68 \\
    & Firecrawl Extract & 1319 &\textbf{45.42}&47.94&87.92&\underline{11.85}&70.47&52.72 \\
    & ReaderLM-v2 & 4000 &26.99&44.35&76.48&7.94&78.42&46.84 \\ \cmidrule{2-9} \rowcolor{lightgray}
   \cellcolor{white} & IndexLM-0.6B & 1966 &40.37&\underline{51.51}&87.76&10.56&\underline{83.31}&\underline{54.70} \\ \rowcolor{lightgray}
   \cellcolor{white} & IndexLM-1.7B & 2043 &41.43&51.42&\underline{88.92}&10.58&\textbf{84.70}&\textbf{55.41} \\ \rowcolor{lightgray}
   \cellcolor{white} & IndexLM-4B & 1928 &\underline{41.56}&\textbf{52.77}&\textbf{89.01}&10.94&82.75&\textbf{55.41} \\ \midrule
    \multirow{10}{*}{\textit{Gemma3-27B-it}}  
    & HTML (raw) & 4096 & 22.01 & 24.91 & 59.98 & 4.20 & 51.19 & 32.46 \\
    & Markdown (raw)     & 4096 & 36.21    & 40.42 & 79.32 & 8.24    & 68.52        & 46.54 \\
    & Chunk-Rerank & 4094 &38.39 & 49.65 & 77.64 & 12.28 &78.98 & 50.79 \\
    & HtmlRAG       & 3562       & 40.87    & 46.99 & 83.72 & 10.94   & 72.37        & 50.98 \\
    & Qwen3-4B + prompt  & 2563       & 43.68    & 48.02 & 85.96 & \underline{15.23}   & 77.85   & 54.15 \\
    & Firecrawl Extract   & 1319   & \textbf{47.67}    & 47.95 & 87.92 & \textbf{20.50}   & 71.12        & 55.03 \\
    & ReaderLM-v2 & 4000 & 30.59 & 43.81 & 79.74 & 8.75 & 75.77 & 47.73 \\ \cmidrule{2-9} \rowcolor{lightgray} 
   \cellcolor{white}  & IndexLM-0.6B   & 1966       & \underline{47.63}    & 52.59 & \underline{88.79} & 15.17   & 84.00   & \underline{57.64} \\ \rowcolor{lightgray}
   \cellcolor{white}  & IndexLM-1.7B   & 2043       & 45.87    & \textbf{54.06} & \textbf{89.02} & 14.32   & \underline{84.21}        & 57.50 \\ \rowcolor{lightgray}
   \cellcolor{white}  & IndexLM-4B     & 1928       & 46.83    & \underline{53.10} & 88.75 & 14.93   & \textbf{86.07}        & \textbf{57.94} \\
    \hline\hline
    \end{tabular}
    \caption{Main results 1: The performance of the extraction model when it serves as a post-retrieval processing component within an RAG QA system. The metric in the table is F1, multiplied by 100 for better display. 'Avg. Tokens' represents the average number of tokens in the final extracted content per question.}
    \label{results1}
\end{table*}

\begin{table*}[htbp]
    \centering
    \setlength{\tabcolsep}{8.5pt}
    \definecolor{lightgray}{gray}{0.96}
    \begin{tabular}{l|>{\columncolor{lightgray}}ccc>{\columncolor{lightgray}}c|>{\columncolor{lightgray}}ccc>{\columncolor{lightgray}}c}
    \hline\hline
    & \multicolumn{4}{c|}{\textit{Main Content Extraction (ME)}} & \multicolumn{4}{c}{\textit{Query-relevant Extraction (QE)}} \\  \cmidrule{2-9}
    \textbf{Method} & \textbf{F1} & \textbf{Precision} & \textbf{Recall} & \textbf{Latency (s) $\downarrow$} & \textbf{F1} & \textbf{Precision} & \textbf{Recall} & \textbf{Latency (s) $\downarrow$} \\ \hline
    HTML (raw)                  & 15.20 & 9.42 & 83.42 & - & - & - & - & - \\
    Markdown (raw)              & 46.07 & 33.93 & \textbf{93.72} & - & - & - & - & - \\
    Chunk-Rerank                & - & - & - & - & 2.83 & 1.46 & \textbf{83.78} & - \\ \midrule
    HtmlRAG                     & 48.65 & 40.57 & 77.52 & 7.12 ($20.3\times$) & 8.83 & 6.95 & 15.98 & 14.62 ($10.5\times$) \\
    Qwen3-4B + prompt           & 71.95 & 80.76 & 69.91 & 17.35 ($49.6\times$) & 26.65 & 27.90 & 31.51 & 11.04 ($7.9\times$) \\
    Firecrawl Extract           & - & - & - & - & 29.48 & \textbf{48.31} & 25.48 & 11.33 ($8.2\times$) \\
    ReaderLM-v2                 & 68.89 & 66.85 & 81.30 & 11.76 ($33.6\times$) & 13.31 & 8.82 & \underline{59.50} & 97.52 ($70.2\times$) \\ \midrule 
    IndexLM-0.6B & \underline{83.38} & \underline{85.28} & 84.63 & \textbf{0.35 ($1.0\times$)} & 28.64 & 33.34 & 37.46 & \textbf{1.39 ($1.0\times$)} \\
    IndexLM-1.7B & 81.78 & 84.16 & 83.44 & \underline{0.42} ($1.2\times$) & \textbf{32} & \underline{37.46} & 39.02 & \underline{1.69} ($1.2\times$) \\
    IndexLM-4B   & \textbf{87.40} & \textbf{85.80} & \underline{92.46} & 0.81 ($2.3\times$) & \underline{31.69} & 37.18 & 39.58 & 3.36 ($2.4\times$) \\
    \hline\hline
    \end{tabular}
    \caption{Main results 2: Directly evaluating by calculating the match rate between its output and the target content. The first four columns are for the 'main content extraction' task, and the last four columns represent the 'query-relevant extraction' task. `Latency' refers to the average time taken to retrieve a webpage.}
    \label{results2}
\end{table*}

\section{Experiments}

\subsection{Testing Datasets and Evaluation Settings}
We utilize two main paradigms to evaluate our extraction method. The first is a downstream task, where we integrate it into an RAG system to see its effect. The second is a direct evaluation, where we measure how well the extracted content matches the ground truth.

\subsubsection{Downstream Task Evaluation: RAG QA} 
First, we evaluated the performance of different extraction methods as post-retrieval processing components within an RAG QA system. In this setting, each question corresponds to about 30 webpages, and the objective of the extraction model is to extract information relevant to the question from these pages.
We select the multi-hop QA datasets HotpotQA \cite{yang2018hotpotqa} and MuSiQue \cite{trivedi-etal-2022-musique}, and the single-hop QA datasets Natural Questions (NQ) and \cite{kwiatkowski2019nq} and TriviaQA (TQ) \cite{joshi-etal-2017-triviaqa}, drawn from the HtmlRAG \cite{htmlrag} test collection. Furthermore, as described in Section \ref{traindata}, a subset of the MultiHopRAG \cite{tang2024multihoprag} data is also incorporated to broaden the distribution of webpages in the test set. Test set statistics are shown in Table \ref{stat}.

We employ Qwen3-4B \cite{yang2025qwen3technicalreport} and Gemma-3-27B-it \cite{gemmateam2025gemma3technicalreport} as the answer models within the RAG system. Given the extraction outputs from each extraction model, these answer models generate responses to questions in the test set. We compute the F1 score between the answer model’s response and the gold answer as the evaluation metric. If a question corresponds to multiple answers, we compute the F1 score between the response and each gold answer, and take the maximum as the final score.

Because a single question is linked to a large number of webpages, potentially exceeding the context limit, we use Qwen-Embedding-0.6B \cite{zhang2025qwen3embeddingadvancingtext} to compute the similarity between each webpage’s extracted content and the question, and use similarity for reranking. At inference time, we insert the extracted content into the answer model’s context in descending order of rank until reaching the maximum context limit. Following HtmlRAG, we report results with a 4K token context as the main setting, and also conduct further experiments with context windows ranging from 0.5K to 32K.

\subsubsection{Direct Evaluation of Extraction Quality}
To more directly assess the performance of the extraction method, we calculate the match rate between the model's output and the target content.
On one hand, we utilize 500 instances randomly sampled from the original training sets of HotpotQA and Musique, for which we retrieve the corresponding Wikipedia pages. These will serve as the experimental data for the `Query-relevant Extraction' (QE) section in Table \ref{results2}. Each webpage in this collection contains the key information required to answer the related question, which allows us to test whether the main content extracted by the model preserves this critical information. We compute F1, precision, and recall between the main content extracted by the different models and the critical information as the metrics (based on tokens).

On the other hand, to more directly evaluate the performance of the main content extraction (ME), we establish detailed annotation guidelines and instruct five human annotators to identify the main content for 100 webpages manually. These pages are sampled from the Common Crawl top 500 domains and the query-relevant test set. 
These webpages are segmented into blocks and indexed using the method described in Section~\ref{index}. Human annotators label the main content by specifying the index ranges corresponding to the webpage’s main content. To ensure the quality of the test set, we calculate the inter-annotator agreement among the five annotators and only retain data points where the pairwise agreement between any two annotators exceeds 80\%. This process yielded a final dataset of 62 items with an average agreement of 91.08\%. Meanwhile, an index is included in the final intervals only if it appears in at least three of the five annotators’ labeled ranges. The final intervals are then converted into text using the method in Section~\ref{postprocessing}. These data will serve as the experimental data for the 'Main Content Extraction' section in Table \ref{results2}. More annotation information can be found in Appendix \ref{annot}.

\subsection{Baselines}
We compare Index-based Web Content Extraction with the following methods:
(1) A simple rule-based approach: we test two strategies—(i) filtering noisy text directly from HTML using cleaning rules (HTML (raw)), and (ii) applying rules to convert the cleaned HTML into Markdown format (Markdown (raw)).
(2) The chunk–rerank method commonly used in RAG: based on the Markdown converted from cleaned HTML, we segment the content into chunks, compute chunk embeddings using Qwen3-embedding-0.6B, and rerank the chunks by similarity.
(3) LLM direct inference: directly extract content from HTML using a prompt + Qwen3-4B.
(4) Specialized web extraction methods, including ReaderLM-v2, HtmlRAG, and the closed-source Firecrawl Extract (due to budget constraints, experiments are conducted on only 10\% of sampled data).

\subsection{Main Results}
\subsubsection{Performance in RAG QA System.}
According to Table \ref{results1}, it is evident that our IndexLM achieves the best average score and matches or outperforms the baselines on every dataset. Meanwhile, the specially trained IndexLM, even with only 0.6B parameters, performs well, which can further improve extraction speed and optimize storage space. As for other baselines, using a proprietary webpage extraction method does improve extraction performance, especially for tasks involving a large number of webpages (e.g., HotpotQA and Musique).

\subsubsection{Performance in Direct Evaluation.} 
Referring to Table \ref{results2}, our index-based extraction maintains high recall while preserving extraction precision. For some methods, the corresponding metrics are missing because they do not support query-relevant or main content extraction. Combined with the results in Table \ref{results1}, we observe that ReaderLM-v2 is better suited for main content extraction. When extracting based on a query, it tends to pull in excessive additional text, leading to high recall but low precision, and its extraction latency is also relatively high. HtmlRAG, on the other hand, requires a preset extraction length, whereas the length of query-relevant content on real webpages is not known in advance. As a result, compared to its score within the RAG QA system, it performs worse on when directly evaluating the extracted output.

\begin{table}[t]
    \centering
    \begin{tabular}{l|ccc}
        \hline\hline
        \textbf{Model} & \textbf{Avg. QA F1} & \textbf{ME F1} & \textbf{QE F1} \\
        \hline
        IndexLM-4B                          & 55.41 & \textbf{87.40} & \textbf{31.69} \\
        \quad\,-- \textit{main content data}       &  \textbf{55.71}  & 70.15 & 30.80 \\
        \quad\,-- \textit{query-relevant data}       & 49.04 & 84.74 & 7.89  \\
        Qwen3-4B + IWE                   & 52.75 & 82.13 & 25.53 \\
        \hline\hline
    \end{tabular}
    \caption{Ablation results. `Avg. QA F1' corresponds to the Average column in Table \ref{results1}, while `ME F1' and `QE F1' are the two F1 columns in Table \ref{results2}. The two middle rows indicate that IndexLM is trained using data from only one task. `Qwen3-4B + IWE' means that within our pipeline, we directly use the original LLM to extract the index instead of IndexLM.}
    \label{ablation}
\end{table}

\begin{figure}[t]
\centering
\includegraphics[width=1\columnwidth]{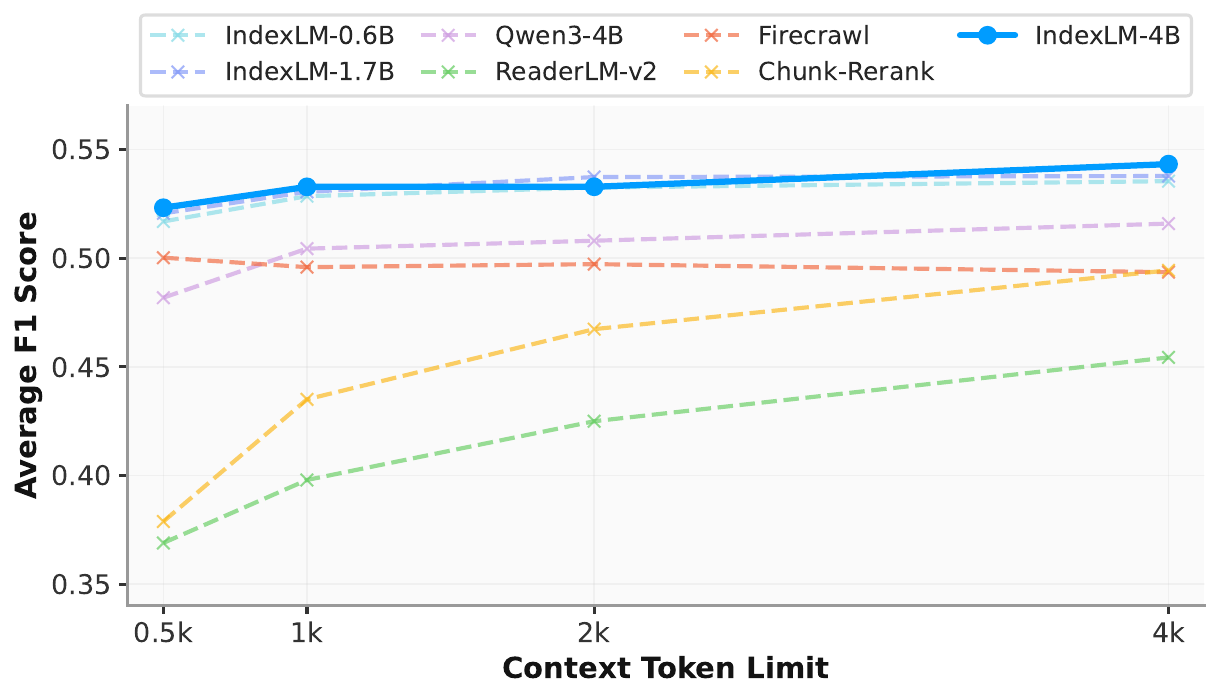}
\caption{Our method consistently outperforms previous works across all context length limits. The curve's stability between 0.5K and 4K suggests that the query-relevant information for most queries is under 512 tokens, and that our approach is able to extract it precisely.}
\label{context1}
\end{figure}

\begin{figure}[t]
\centering
\includegraphics[width=1\columnwidth]{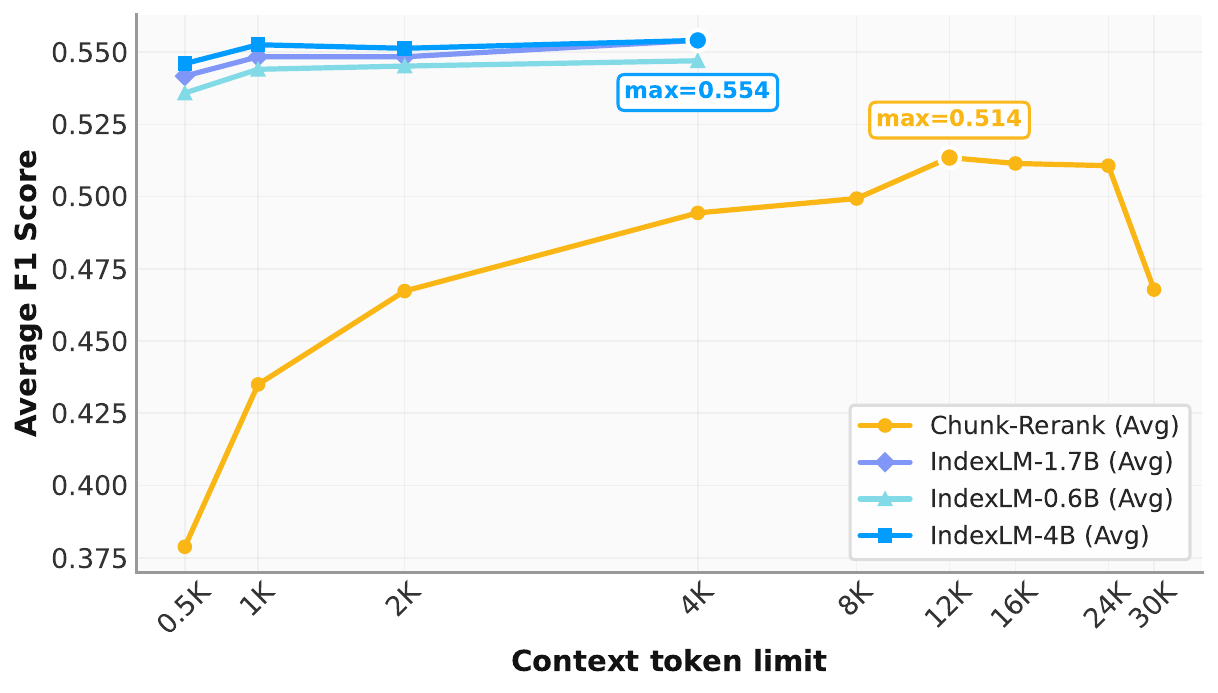}
\caption{The peak score of the traditional RAG Chunk-rerank approach, even with an unlimited context length, is surpassed by our method's score under 0.5K token limit.}
\label{context2}
\end{figure}

\subsection{Analysis}
\subsubsection{Extraction Speed}
As presented in Table \ref{results2}, we measure the average end-to-end latency, from receiving a webpage to completing extraction, for each method. On a standardized Nvidia A800 setup, index-based extraction is up to 10× faster than generative approaches. For main content extraction, where more content must be extracted from each webpage, our method’s speed advantage is even more pronounced.
Figure \ref{pace} further illustrates this point: Our IndexLM is both more effective and efficient compared to generative extraction models. This is primarily because the number of tokens directly output by the model is significantly lower than with other models. For instance, in the main content extraction task, IndexLM-4B outputs an average of 25 tokens, whereas ReaderLM-v2 outputs 2,308 and Qwen3-4B with prompt outputs 1570. Moreover, the 0.6B and 1.7B versions offer even greater extraction speed by sacrificing a small amount of accuracy. The data used in Figure \ref{pace} is the average of the data corresponding to the two tasks in Table \ref{results2}.

\subsubsection{Ablation}
We conduct several ablation experiments. From Table \ref{ablation}, training IndexLM with only query-relevant data yields normal performance on the query-relevant task, with some degradation on main content extraction, though it remains workable. In contrast, training with only main content data almost fails on the query-relevant task.
Replacing IndexLM in our Index-based Web Content Extraction framework with the original, untrained Qwen3-4B leads to a performance drop compared to IndexLM-4B, but the decline is marginal. This confirms IndexLM’s effectiveness and also shows our framework’s inherent effectiveness and generality, as it performs reasonably well even without a specialized model.

\subsubsection{Influence of Context Length Limit}
In Main Results 1, we set the answer model’s context limit to 4K. We also conducted experiments with context limits of 0.5K–2K. As shown in Figure \ref{context1}, where the y-axis represents the mean score of each method across the five datasets, our method outperforms other baselines regardless of the context limit. Meanwhile, the extraction performance changes little with context length because: first, for many queries, the total number of relevant content tokens in the corresponding webpages is smaller than 4K, as indicated in the “Avg. Tokens” column of Table \ref{results1}. Second, our method can precisely extract while avoiding irrelevant information, thereby reducing the LLM’s context load.

This raises a new question. As seen in Figure \ref{context1}, the Chunk-Rerank approach commonly used in traditional RAG also improves as the context limit increases. Will it surpass our model when the context limit is longer? We further conducted experiments with context limits from 8K to 30K. As shown in Figure \ref{context2}, when the context limit increases, the performance of Chunk-Rerank does improve further and peaks at 12K. But as noise in the extracted content grows, its performance declines. However, even at its peak, it is still below the lowest result of our index-based extraction, so this new concern no longer exists.

\section{Conclusion}
In this paper, we introduce Index-based Web Content Extraction, a new method that reframes extraction from token-by-token content generation to index prediction. It can be used to address the problem that today’s RAG systems and web agents need to read massive volumes of web pages that have large token budgets and low signal density. We validate our method with multiple experiments, showing it is both effective and efficient, outperforming different baselines in accuracy and speed.
Our method also has limitations. For example, we train our extraction model using SFT. In the future, reinforcement learning methods could be used to further enhance the model’s capabilities. In addition, our index-based extraction approach can be extended to other domains beyond web content, and we hope that this will inspire future work.


\bibliographystyle{ACM-Reference-Format}
\bibliography{sample-sigconf-authordraft}

\appendix

\section{Method Details} \label{method_details}

\begin{figure*}[t]
\centering
\includegraphics[width=1\linewidth]{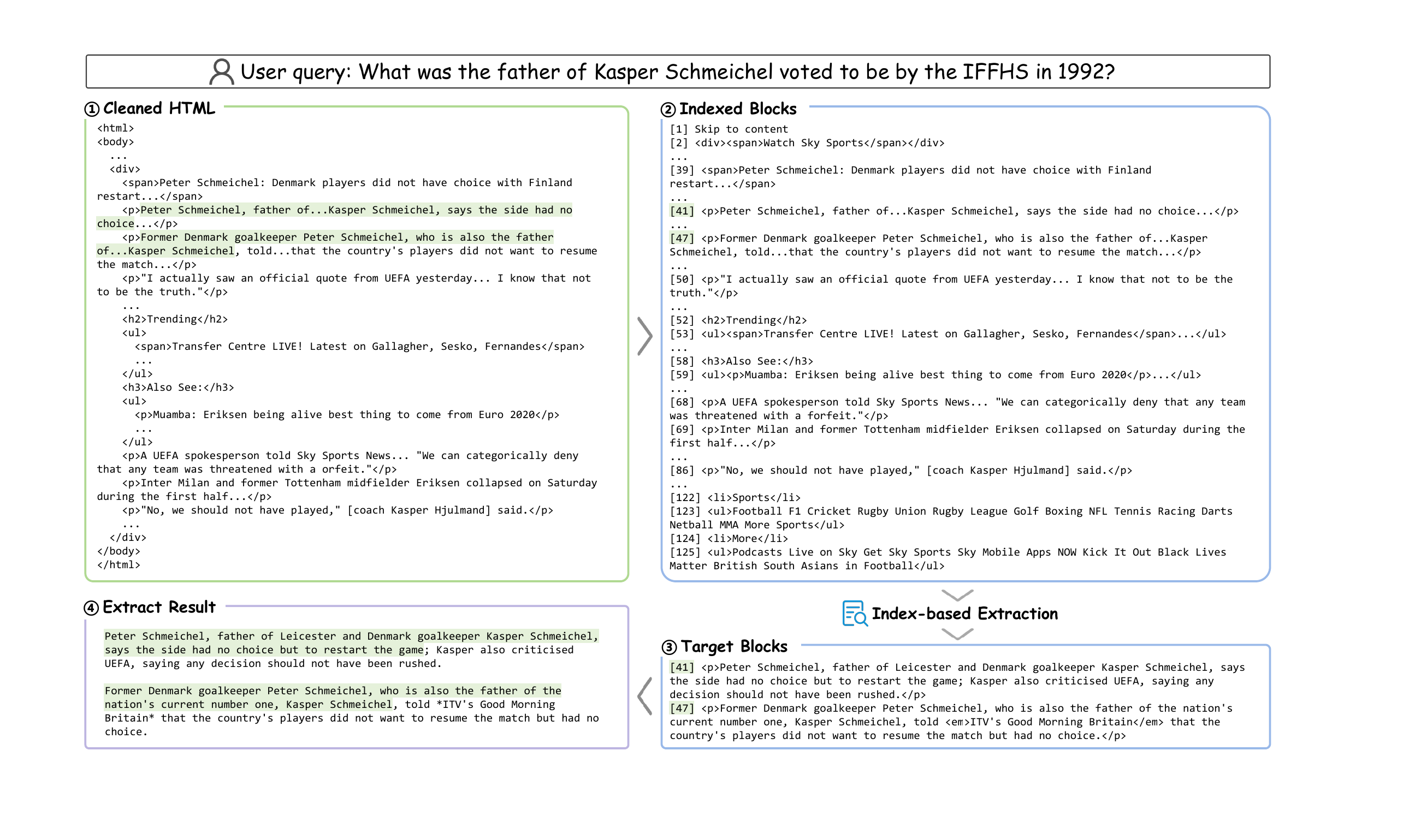}
\caption{A more detailed example from HotpotQA for Index-based Web Content Extraction.}
\label{example2}
\end{figure*}

\begin{figure}[t]
\centering
\includegraphics[width=1\columnwidth]{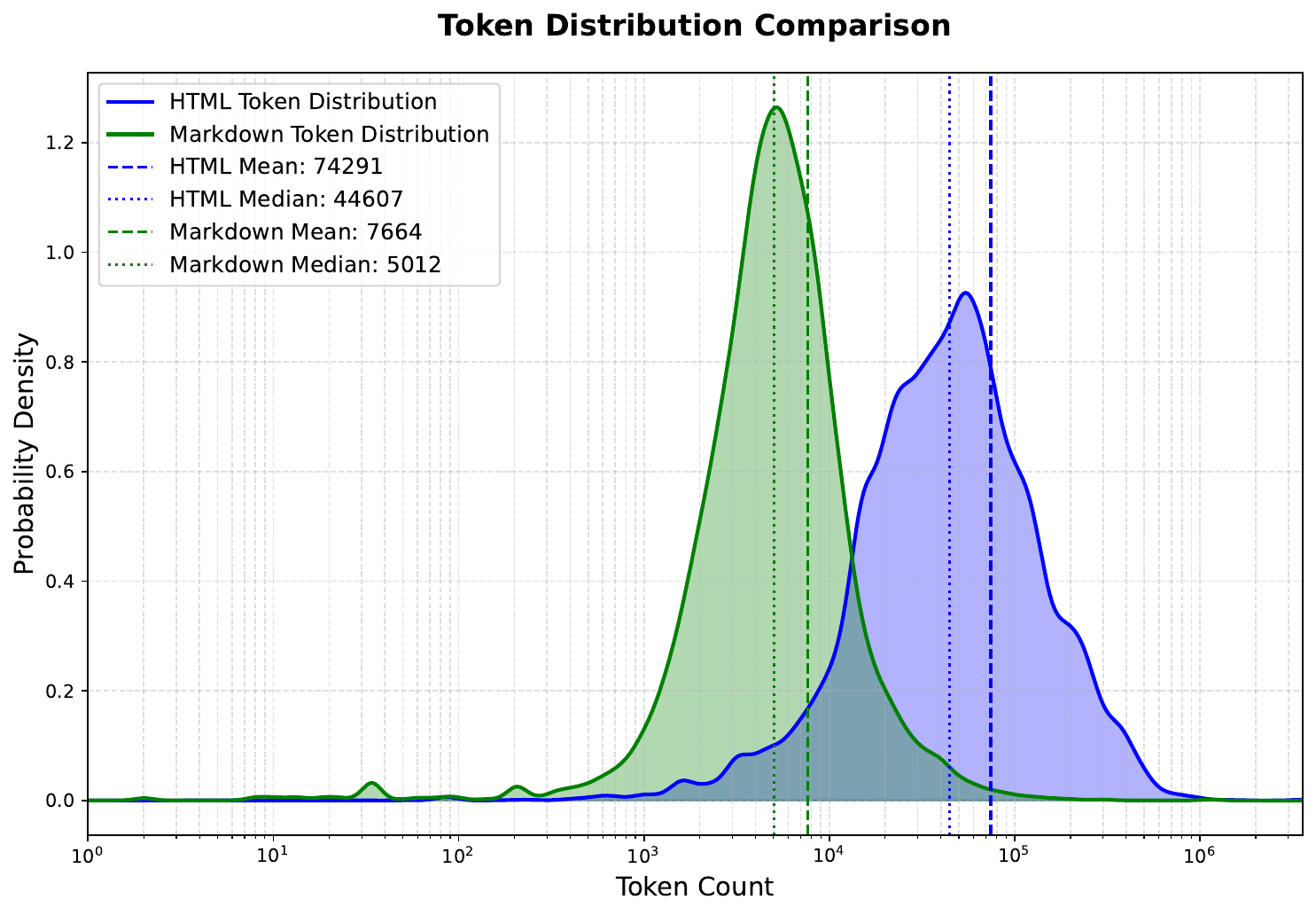}
\caption{Distribution of token counts for all web pages in the test set. Blue represents the total token count of the original HTML, and green represents the total token count after directly converting the original HTML to Markdown.}
\label{dis}
\end{figure}

\subsection{Statistics} \label{example}
We conduct a length distribution analysis on all web pages used in our experiments (see Figure \ref{dis}).
The figure reveals that unprocessed web page HTML averages over 70K tokens, and even when using a mature tool to convert it to Markdown, the average length remains nearly 8K.
If an LLM is limited to a 32K context window, it can process a maximum of only four web pages. This is clearly insufficient for a web agent and further underscores the necessity of effective web content extraction.

\subsection{Example} \label{example}
Due to space limitations in the main paper, we presented a simplified example. Here, we provide a more detailed, real-world extraction case (see Figure \ref{example2}).

\subsection{Content Segmentation and Indexing} \label{segment}
\paragraph{Block-level elements}
\textbf{(1)} If a parent element contains direct text (defined as text not encapsulated within a child block-level element), the parent’s textual content and its inline element text are consolidated into a single segment; each child block-level element is then recursively processed, with its resulting segments appended in sequential order.
\textbf{(2)} If a parent element lacks direct text but contains child block-level elements, it will be merged with the first non-empty child segment during the post-processing stage. Specifically, a new segment is formed by enclosing the first child block-level element within the parent's start and end tags (e.g., \texttt{<div><p>text</p></div>}). The remaining child block-level elements are then output sequentially as independent segments.

\paragraph{Inline-level elements}
\textbf{(1)} Images are preserved only if they are associated with a textual caption. They are subsequently reformatted into a standardized string: \texttt{<img>image: [link], caption: [text]</img>}. Images lacking a caption are discarded to reduce noise from non-semantic visual elements.
\textbf{(2)} For inline elements that primarily denote formatting, such as \texttt{<b>}, \texttt{<i>}, \texttt{<u>}, as well as \texttt{<br>} and \texttt{<code>}, their tags are retained. For all other inline elements, only their textual content is preserved while the tags themselves are stripped. The original sequence of all inline content within a block-level element is maintained after processing.
\textbf{(3)} In the case of hyperlinks (\texttt{<a>}), only the visible anchor text is retained. The underlying URL is discarded to eliminate extraneous information.

Given that textual content within an element can be excessively long, we define a maximum length limit $L_{max, block}$. To avoid disrupting the text's structure, any block exceeding this limit will be split. 
If a resulting segment remains oversized, it is recursively split by sentences and then by words. Each fragment is then wrapped in its original parent HTML tag and assigned a triplet of attributes for traceability: \texttt{split-id} (source block identifier), \texttt{split-part} (fragment sequence number, starting from 1), and \texttt{split-total} (total fragment count). This mechanism allows for the accurate reconstruction of the original content.

\begin{figure*}[htbp]
\centering
\includegraphics[width=1\linewidth]{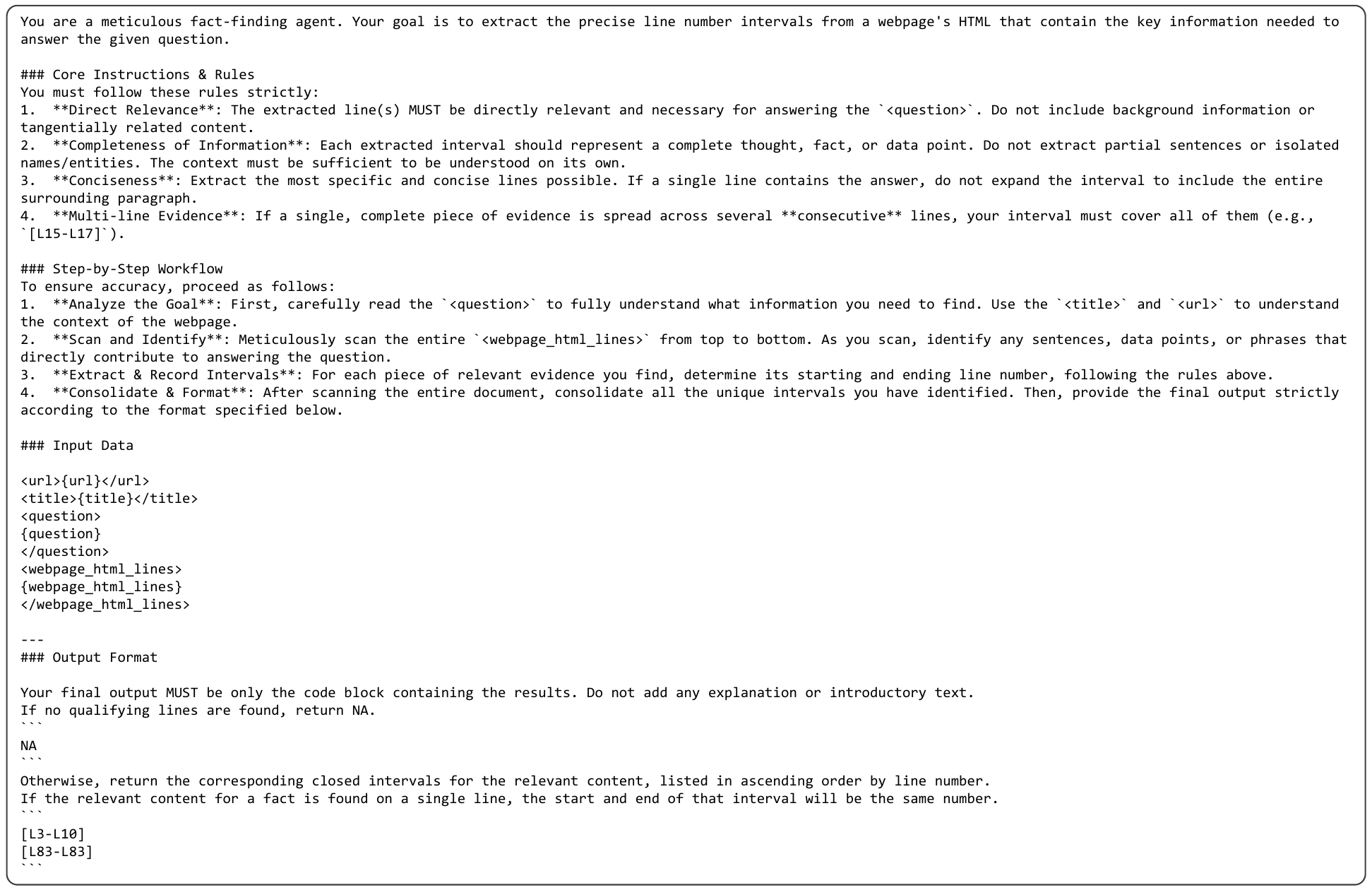}
\caption{The prompt of Index-based Web Content Extraction}
\label{prompt}
\end{figure*}

\subsection{Post-processing} \label{post}
The primary objective of the post-processing stage is to reassemble the blocks into a single HTML document. This approach allows for the use of mature tools that can convert HTML into various other text formats, such as the commonly used Markdown format \cite{JinaAI2024ReaderLM}.

The first step is to concatenate content blocks that were split from a single element due to length constraints, based on their attribute triplets. For the same `split-id`, the text within the tags is concatenated in ascending order of the `split-part` value, after which the block-splitting tags are removed from the element.

The second step is the reconstruction process. This step primarily focuses on the parent-child structure of block-level elements, which was disrupted during the splitting phase. We designed a heuristic method to restore common parent-child structures in the DOM tree that may affect text formatting, such as `<p>`, heading elements (`<h1>`-`<h6>`), `<ul>`, `<li>`, and `<table>`. When a pattern of a "parent element wrapping the first child segment" is detected, it is identified as a collapsed empty parent container. Then, existing heuristics are used to continuously gather subsequent child segments, which are finally enclosed within the parent container as a single output. For example, upon detecting a `<ul>` followed by an `<li>`, it is identified as the start of an unordered list. The process then automatically detects all subsequent `<li>` elements until the list concludes and places all of them within the `<ul>` container.

As for inline elements, images will be reformatted from our structured string back to their original format, containing only the image link and its caption. Since other format-related inline elements already have their tags preserved, no additional operations are performed on them.

\section{Experiment Details}

\subsection{Human Annotation} \label{annot}
We built our evaluation test set for main content extraction through manual human effort. First, we developed a web-based annotation platform so that annotators could conveniently perform their labeling tasks. Subsequently, the annotators are trained to ensure they fully understood the motivation behind the task, the detailed rules, and the definition of a webpage's main content.

The human annotation process required a meticulous examination of each webpage. On average, each expert annotator spent approximately 10 hours to annotate 100 webpages. This process culminated in a total of 50 person-hours of human annotation across all tasks and annotators. We also provided the annotators with reasonable compensation for their work.

\subsection{Prompt} \label{prompt}
The prompt of our Index-based Web Content Extraction is displayed in Figure \ref{prompt}. Constrained by space, we only display the prompt for the query-relevant task here. The prompt used for main content extraction was slightly adjusted from this prompt.

\end{document}